\documentclass[12pt]{article} 
\usepackage[bookmarks=true,bookmarksnumbered=true,setpagesize=false]{hyperref} 
\usepackage{amsmath,amssymb,bm,epsfig,color,graphicx,appendix,comment}
\renewcommand{\thefootnote}{\fnsymbol{footnote}}
\usepackage[normalem]{ulem}

\begin{document}

\title{
\begin{flushright}
\begin{minipage}{0.2\linewidth}
\normalsize
\end{minipage}
\end{flushright}
{\Large \bf 
Cosmological Perturbations in Palatini Formalism
\\*[20pt]}}

\author{
Mio Kubota$^{1}$\footnote{
E-mail address: mio.kubota@hep.phys.ocha.ac.jp},
Kin-ya Oda$^{2}$\footnote{
E-mail address: odakin@phys.sci.osaka-u.ac.jp},
Keigo Shimada$^{3}$\footnote{
E-mail address: shimada.k.ah@m.titech.ac.jp}, and\ 
Masahide Yamaguchi$^{3}$\footnote{
E-mail address: gucci@phys.titech.ac.jp}\\*[20pt]
$^1${\it \normalsize
Department of Physics, Ochanomizu University, Tokyo 112-8610, Japan} \\
$^2${\it \normalsize
Department of Physics, Osaka University, Osaka 560-0043, Japan} \\
$^3${\it \normalsize
Department of Physics, Tokyo Institute of Technology,
Tokyo 152-8551, Japan} \\*[50pt]
}

\date{
\centerline{\small \bf Abstract}
\begin{minipage}{0.9\linewidth}
\medskip
\medskip 
\small
\quad

We investigate cosmological perturbations of scalar-tensor theories in
Palatini formalism. First we introduce an action where the Ricci scalar
is conformally coupled to a function of a scalar field and its kinetic
term and there is also a k-essence term consisting of the scalar
and its kinetic term. This action has three frames that are equivalent
to one another: the original Jordan frame, the Einstein frame where the
metric is redefined, and the Riemann frame where the connection is
redefined. For the first time in the literature, we calculate the
quadratic action and the sound speed of scalar and tensor perturbations
in three different frames and show explicitly that they
coincide. Furthermore, we show that for such action the sound speed of
gravitational waves is unity. Thus, this model serves as dark energy as
well as an inflaton even though the presence of the dependence of the kinetic term of a scalar field in the non-minimal coupling, different from the case in metric formalism.  We then proceed to construct the L3
action called Galileon terms in Palatini formalism and compute its
perturbations. We found that there are essentially 10 different
(inequivalent) definitions in Palatini formalism for a given Galileon
term in metric formalism. We also see that, in general, the L3 terms
have a ghost due to Ostrogradsky instability and the sound speed of
gravitational waves could potentially deviate from unity, in sharp
contrast with the case of metric formalism. Interestingly, once we
eliminate such a ghost, the sound speed of gravitational waves also
becomes unity. Thus, the ghost-free L3 terms in Palatini formalism can
still serve as dark energy as well as an inflaton, like the case in metric formalism.
\end{minipage}
}

\maketitle
\tableofcontents
\clearpage

\renewcommand{\thefootnote}{\arabic{footnote}}
\setcounter{footnote}{0}

\def\eqn#1{\begin{eqnarray}#1\end{eqnarray}}
\def\eqm#1{\begin{eqnarray*}#1\end{eqnarray*}}
\def\l{\left}
\def\r{\right}
\def\={&=&}
\def\nn{\nonumber}
\section{Introduction}

There have been long attempts to extend general relativity since
the Brans-Dicke theory~\cite{Brans:1961sx}. Such scalar-tensor theories have recently received renewed attention because a scalar dynamical degree of freedom may possibly be the cause of dark energy and well be of inflation~\cite{Starobinsky:1980te,Sato:1980yn,Guth:1980zm}, both of which enjoy strong observational support; see Refs.~\cite{Perlmutter:1998np,Riess:1998cb} and e.g.\ Ref.~\cite{Akrami:2018odb}, respectively.
In a scalar-tensor theory, a scalar field does not necessarily couple to gravity (a tensor field) minimally but non-minimally.  One
major example is the Higgs inflation
\cite{Salopek:1988qh,CervantesCota:1995tz,Bezrukov:2007ep}, which was
originally proposed by introducing such a non-minimal coupling of a
scalar field to the Ricci scalar \cite{Futamase:1987ua}.

When one introduces such a non-minimal coupling between the scalar and tensor fields, two different approaches are commonly considered in the literature: one is called the metric formalism and the other the Palatini formalism.\footnote{It has been reported~\cite{Ferrarisetal}
that the Palatini formalism is not invented by Palatini, but by
Einstein. Here we just follow the convention for the naming. Instead,
one may call it the metric-affine theory~\cite{Sotiriou:2006mu}. Later
in Ref.~\cite{Sotiriou:2008rp}, the authors made a distinction such that the Palatini and metric-affine theories refer to the cases in which the matter action does and does not contain the independent connection, respectively.}  In
the former formalism, an affine connection is not an independent
variable in the action but given solely by the metric, that is, one {\it
a priori} decides to use the Levi-Civita connection. On the other hand,
in the latter formalism, a connection is regarded as an independent
variable in a Jordan-frame action, and is fixed {\it or
solved} through Euler-Lagrange equations, which are given by taking the
variations of the action with respect to not only the metric (and
matter) but also the connection.

In Einstein gravity (in the Einstein frame), where the action
solely consists of the Einstein-Hilbert action (with the connection) and
a matter action {\it without} the independent connection, the connection
reduces to the Levi-Civita connection even in the Palatini
formalism. Thus, both formalisms lead to the same theory in Einstein
gravity. However, generally speaking, the metric and Palatini formalisms
give different theories when one starts from a given Jordan-frame
action, that is, different predictions for the same Jordan-frame action
in a scalar-tensor theory with a non-minimal coupling. It is known that
in the context of theories with purely metric and connection, the only
theories that are equivalent between the metric formalism and Palatini
formalism are theories that exhibit the same structure as Lovelock
gravity~\cite{Exirifard:2007da,Borunda:2008kf}.  Whether there exists
such equivalence in scalar-tensor theories, if at all, is still unknown to our best knowledge. Furthermore, some difficulties were pointed out
in a class of gravity theories like $f(R)$ gravity in Palatini formalism
(see e.g.\ Refs.~\cite{Iglesias:2007nv,Sotiriou:2008rp}; see also
Appendix B in Ref.~\cite{Jinno:2019und}). They {\it may} be resolved by
extending Palatini formalism to the metric-affine formalism, where the
coupling of matter to gravity can include not only a metric but also a
connection.

The local-Lorentz (LL) symmetry is indispensable just to {\it define} a
spinor field on curved background~\cite{Utiyama:1956sy}, namely to
define matter in our universe.  Given the tremendous success of gauge theory in establishing the Standard Model of particle physics, it
is tempting to introduce a LL gauge field as an independent dynamical
variable; see e.g.\ Ref.~\cite{Matsuzaki:2020qzf}.  A variation with
respect to the LL gauge field is equivalent to the variation with
respect to the connection. In this regard, one might find it more
natural to work in the Palatini formalism.

At present, since Einstein gravity fits observations very well, one cannot judge which
formalism is adopted by nature. Such judgment can be done only through
comparing theoretical predictions in both formalisms with the
observations that may arise beyond General Relativity.
 
For this purpose, people have recently  constructed viable scalar-tensor theories and made theoretical predictions for inflation in Palatini formalism and investigated the difference between
those in metric and Palatini
formalisms~\cite{Lindstrom:1975ry,Lindstrom:1976pq,Bauer:2008zj,Li:2012cc,Markkanen:2017tun,Jarv:2017azx,Aoki:2018lwx,Kozak:2018vlp,Shimada:2018lnm,Jinno:2018jei,Aoki:2019rvi,Helpin:2019kcq,Helpin:2019vrv,Mikura:2020qhc};
see Ref.~\cite{Tenkanen:2020dge} for a recent review. One promising
example is Higgs inflation in Palatini
formalism~\cite{Rasanen:2017ivk,Rasanen:2018ihz,Jinno:2019und,Tenkanen:2019jiq}. In
the case of Higgs inflation, where the Higgs field directly couples to
the Ricci scalar, one can always go into the Einstein frame through a
conformal transformation. Of course, the correspondence between the Einstein and Jordan frames is different in metric and Palatini formalisms.
That is, when we start from an original action in the Jordan frame in general, the corresponding action in the Einstein frame is different between these formalisms. By the use of the standard formulae for
inflationary predictions in the Einstein frame, one can make different
predictions for metric and Palatini formalisms
\cite{Rasanen:2017ivk,Jinno:2019und,Tenkanen:2019jiq}.\footnote{The
equivalence between the Einstein and Jordan frames through conformal or
disformal transformation is discussed e.g. in
Refs.~\cite{Catena:2006bd,Deruelle:2010ht,Chiba:2013mha,Domenech:2015hka,Chiba:2020mte}}

However, in a wider class of scalar-tensor theories, the Einstein frame
does not necessarily exist.  Besides, even in the case of Higgs
inflation, there is a subtle issue when one considers matter couplings. Even if matter (other than Higgs) minimally couples
to gravity in the Jordan frame, it ``non-minimally" couples to gravity
in the Einstein frame through conformal transformation. Thus, a
scalar-tensor action in the Jordan frame with a minimally coupled matter
action is not equivalent to a corresponding scalar-tensor action in the
Einstein frame with a ``minimally" coupled matter action, whose
difference can be probed by reheating process, for example, which is an
interesting topic though we do not deal with it in this paper. In addition, in Refs.~\cite{Helpin:2019kcq,Helpin:2019vrv}, the
authors extended Horndeski theories into Palatini formalism and
discussed its possible applications.  In such theories that include the
connection through covariant derivatives, the Einstein frame does not
exist.

Therefore, it is quite useful and interesting to discuss cosmological
perturbations directly in the Jordan frame in the context of Palatini formalism without resorting to the conformal
transformation into the Einstein frame. Furthermore, as far as we are aware of, nobody has
yet derived quadratic actions for cosmological scalar and tensor
perturbations in the Jordan frame in the context of Palatini formalism
of scalar-tensor theory.  Such quadratic actions are indispensable for
determining normalizations of the perturbations and manifestly give
their sound speeds. Thus, the main purpose of this paper is to derive
quadratic actions for cosmological scalar and tensor perturbations and
to discuss their properties based on the quadratic actions directly in
the Jordan frame as well as in the Einstein frame where the
metric is redefined, and in the Riemann frame where the connection is
redefined, in the context of Palatini formalism. 

As an example of
the absence of the Einstein frame, we consider so-called the Galileon
(Kinetic Gravity Braiding) term
\cite{Nicolis:2008in,Deffayet:2010qz,Kobayashi:2010cm} as well as the
non-minimal coupling of the Ricci scalar and the k-essence term \cite{ArmendarizPicon:1999rj,Chiba:1999ka} in our
action. As pointed out in
Refs.~\cite{Aoki:2018lwx,Shimada:2018lnm,Helpin:2019kcq,Aoki:2019rvi,Helpin:2019vrv},
A term corresponding to $\Box\phi$ in metric formalism is not uniquely defined in Palatini
formalism due to the covariant derivative not being compatible with the
metric, i.e.\ the presence of non-metricity. We have found that there are
{essentially 10} different (inequivalent) definitions in Palatini
formalism for such a term, and have included all of them in our action. In the
case of metric formalism, it is known that the sound speed of tensor perturbations
(gravitational waves) is still unity even if one includes the Galileon
term in an action~\cite{Kobayashi:2011nu}. Therefore it can serve as dark
energy even after the observation of GW170817 and GRB170817A, which
strongly constrains the sound speed of gravitational waves
\cite{Monitor:2017mdv}. We are going to address whether the Galileon
term would modify the sound speed of gravitational waves in Palatini
formalism or not, which is crucial for serving as dark energy.

This paper is organized as follows. First in Sec.~\ref{sec:GR} we shall
briefly review Palatini formalism for the case of the Einstein-Hilbert
action and show the equivalence with the case of metric formalism.
In Sec.~\ref{sec:Ricci}, we derive quadratic actions of tensor and
scalar perturbations in three different frames for an action consisting of k-essence action as
well as a non-minimal coupling of the scalar field to Ricci scalar. 
{Though the Jordan frame action can be
recast into that in the Einstein frame through conformal transformation,
it is instructive to demonstrate how to obtain quadratic actions by
perturbing not only a metric and a scalar field but also a connection.
We also introduce a \emph{Riemann frame}, where the dynamics of the metric
and scalar fields is still governed by a Jordan-frame action but is written instead in Riemann geometry, i.e., the connection is replaced, using its equation of motion, by the Levi-Civita plus the matter terms.
We also calculate quadratic actions for scalar and tensor perturbations
in this Riemann frame as well. We discuss their power spectra as well as
their sound speeds in all the three frames separately, and show
their equivalence explicitly.} In Sec.~\ref{sec:Galileon}, we discuss possible
{L3} terms in Palatini formalism, which correspond to the Galileon
term in metric formalism. In Sec.~\ref{sec:quadratic}, we first
derive the equivalent action in Riemann frame and discuss the ghost-free
condition. We also derive quadratic actions of tensor and scalar
perturbations for an action including such Galileon terms and estimate
their power spectra as well as their sound speeds. The final section is
devoted to conclusions and discussions.

\section{Brief Review of General Relativity in Palatini Formalism}
\label{sec:GR}

It is well known that, for the Einstein-Hilbert action, both the metric and Palatini formalisms result in the same equation of motion.
Here, we will review Palatini formalism in the context of General
Relativity and see explicitly that this fact holds.

Firstly, a covariant derivative of a general affine-connection is defined as
\eqn{
{\overset \Gamma\nabla}_\mu A_\nu :=\partial_\mu A_\nu -\Gamma^\lambda{}_{\nu\mu} A_\lambda.
}
In the presence of a metric $g_{\mu\nu}$, this affine connection can be uniquely decomposed as follows,
\eqn{
\Gamma^\lambda_{~\mu\nu} \=\l\{\substack{\lambda\\\mu\nu}\r\}_g+\frac12\l(T_{\mu\nu}{}^{\lambda}+T_{\nu\mu}{}^{\lambda}-T^\lambda{}_{\mu\nu}+Q_{\mu\nu}{}^{\lambda}+Q_{\nu\mu}{}^{\lambda}-Q^\lambda{}_{\mu\nu}\r),
}
where the Levi-Civita connection of the metric $g_{\mu\nu}$, torsion,
and non-metricity are defined respectively as
\eqn{
\l\{\substack{\lambda\\\mu\nu}\r\}_g&:=&\frac12g^{\lambda\sigma}\l(\partial_\mu g_{\nu\sigma}+\partial_\nu g_{\mu\sigma}-\partial_\sigma g_{\mu\nu}\r),\\
T^\lambda{}_{\mu\nu}&:=&\Gamma^\lambda{}_{\nu\mu}-\Gamma^\lambda{}_{\mu\nu},\\
Q_\sigma{}^{\mu\nu}&:=&\overset\Gamma\nabla_\sigma g^{\mu\nu}.
}

With the covariant derivative, one may define its ``field strength'', the Riemann tensor, as,
\eqn{
[{\overset\Gamma\nabla}_\mu,{\overset\Gamma\nabla}_\nu]A^\lambda&=:&{\overset\Gamma R}{}^{\lambda}{}_{\sigma\mu\nu}A^\sigma-T^\sigma{}_{\mu\nu}\overset\Gamma\nabla_\sigma A^\lambda,
}
where the explicit form of the Riemann tensor is
\eqn{
{\overset\Gamma R}{}^{\lambda}{}_{\sigma\mu\nu}&:=&\partial_\mu\Gamma^\lambda{}_{\sigma\nu}-\partial_\nu\Gamma^\lambda{}_{\sigma\mu}+\Gamma^\lambda{}_{\rho\mu}\Gamma^\rho{}_{\sigma\nu}-\Gamma^\lambda{}_{\rho\nu}\Gamma^\rho{}_{\sigma\mu}.
}
The action constructed from this Riemann tensor at its lowest order is none other than the Einstein-Hilbert action
\eqn{
S_{\text{EH}}=\int d^4x \sqrt{-g}{\overset\Gamma R},
}
where the Ricci tensor and scalar are defined as
\eqn{
{\overset\Gamma R}{}_{\mu\nu}&:=&{\overset\Gamma R}{}^{\lambda}{}_{\mu\lambda\nu},\\
{\overset\Gamma R}&:=&g^{\mu\nu}{\overset\Gamma R}{}_{\mu\nu}.
}
Note that the Ricci tensor is neither symmetric nor unique, since one may consider two other contractions of the Riemann tensor, namely the co-Ricci tensor and homothetic tensor. However, such ambiguity of the Ricci tensor is not important for this paper since we only consider the Ricci scalar, which is unique.
Here and after we shall consider the connection to be torsionless just for simplicity.\footnote{In general, a theory computed with a general affine connection and one with a torsionless connection differ~\cite{Olmo:2013lta,Afonso:2017bxr}.  However, under special cases, such as when the action respects a symmetry called projective invariance,  some theories coincide whether torsionlessness is assumed or not. Theories such as the Einstein-Hilbert action and k-essence action} with non-minimal coupled Ricci scalar are such cases.

The equation of motion for the metric results in
\eqn{
\frac{\delta S_{\text{EH}}}{\delta g^{\mu\nu}}=
\overset\Gamma R{}_{(\mu\nu)}-\frac12 \overset{\Gamma}R g_{\mu\nu},
	\label{eq:Einstein}
}
which are none other than the (Palatini) Einstein equations.

Assuming that matter does not couple to the connection and then solving the equation of motion for the (torsionless) connection, we obtain,
\eqn{
\Gamma^\lambda_{~\mu\nu} \=\l\{\substack{\lambda\\\mu\nu}\r\}_g.
}
Therefore, substituting this to the Palatini Einstein equations (\ref{eq:Einstein}), one obtains precisely the usual Einstein equations.  Therefore, the Einstein-Hilbert action for both metric and Palatini formalism computes precisely the same equation of motion. 

In the presence of non-trivial matter, this equivalence is known not to hold. For example, the presence of minimally coupled fermions effectively gives four-point fermion couplings and is known as
the Hehl-Datta Lagrangian~\cite{Hehl:1971qi}.  In the Palatini formalism,
however, it is often assumed that matter $\Psi$ only couples to
the metric\footnote{
In an extended scenario of metric-affine formalism, matter can be considered to be coupled with the connection. Cosmological implications of such models can be seen in, for example, \cite{Obukhov:1993pt,Iosifidis:2020gth,Iosifidis:2020zzp,Iosifidis:2020upr}}with the form,
\eqn{
S=S_{\text{gravity}}+\int d^4x\sqrt{-g}\mathcal L_\text{m}(g_{\mu\nu},\Psi).
}
As a result of this form, matter is conserved under the covariant derivative with respect to Levi-Civita $\overset{g}\nabla$ due to the diffeomorphism invariance of the matter action:
\eqn{
\overset{g}\nabla_\mu T^{\mu\nu}=0,
}
where the energy-momentum tensor is defined as usual as
\eqn{
T_{\mu\nu}:=-\frac2{\sqrt{-g}}\frac{\delta \mathcal L_\text{m}}{\delta g^{\mu\nu}}.
}

Due to the conservation of the energy-momentum tensor with respect to
the Levi-Civita connection, the conservation of curvature perturbations
on superhorizon scales is also assured even in Palatini formalism
unless there are non-adiabatic pressure perturbations. This fact is not
explicitly mentioned in the literature but quite important in
cosmological applications. Furthermore, it is also important to notice
that the particle motions of matter follow (extremal) geodesics and not
of auto-parallels.

\section{K-essence with its non-minimal coupling to Ricci scalar} 
\label{sec:Ricci}

In this section, we will first consider an action that has a conformally coupled Ricci scalar as follows
\begin{equation}
S_4^{~\text{Jordan}}:= \int d^4 x \sqrt{-g} \left[G_4(\phi,X) \overset{\Gamma}{R} + K(\phi,X)  \right],\label{eq:L4}
\end{equation}
with $X:=-\frac12g^{\mu\nu}\partial_\mu \phi\,\partial_\nu\phi$.

In the metric formalism, in order to keep the equation of motion second-order, it is known that one must introduce a ``counter term", namely
\eqn{
\mathcal L_4^{~\text{metric}}=G_4(\phi,X)\overset{g}{R}+G_{4X}\l[\l(\phi_{\mu\nu}\r)^2-(\phi_\mu{}^{\mu})^2\r],
\label{eq:L4metric}
}
where $\phi_\mu:=\partial_\mu\phi$ and $\phi_{\mu\nu}:=\overset{g}{\nabla}_\mu\overset{g}{\nabla}_\nu \phi$. However, for the Palatini Lagrangian (\ref{eq:L4}) this counter term is unnecessary to keep the equation of motion to be second-order. Furthermore, as we will mention in detail later, the covariantization of the counter term in Palatini formalism may make the connection dynamical and introduce new degrees of freedom.

This action (\ref{eq:L4}) can be investigated through three methods. In the first method, as most literature considers, one can conduct a conformal transformation of the metric to the Einstein frame and calculate everything there. This method is useful when there exists an Einstein frame. The second way is to directly calculate within the Jordan frame~(\ref{eq:L4}). Although tedious, this is the most straightforward method. Finally, another less-known method is to solve the connection, which is non-dynamical, and substitute the solution to the action. The resultant (on-shell) action is written fully in terms of Riemann geometry. This results in an action that is neither Einstein nor Jordan. We shall call this frame, where the connection rather than the metric is redefined, the Riemann frame. 

Since all three frames are nothing but (invertible) redefinitions of physical
variables, one expects to see that the results of calculations in
three different frames coincide. We shall see this in the
following sections.

\subsection{Analysis in Einstein frame}

Similarly to the usual case of metric formalism, consider the conformal transformation of the action (\ref{eq:L4}) under
\eqn{
{\tilde g}_{\mu\nu}=G_4 g_{\mu\nu}.
}
The action then becomes
\eqn{
S_4^{~\text{Einstein}}=\int d^4\tilde x \sqrt{-{\tilde g}}\l[{\tilde g}^{\mu\nu}\overset{\Gamma}{R}_{\mu\nu}+\frac{K(\phi,G_4 \tilde X)}{{G_4}^2(\phi,G_4 \tilde X)}\r],
}
where $\tilde X:=-\frac12{\tilde g}^{\mu\nu}\tilde\partial_{{\mu}}\phi\tilde\partial_{{\nu}}\phi=G_4^{-1}X$ in which $\tilde{x}^\mu$ is the coordinate in the Einstein frame; see below. 

This term is none other than the Einstein-Hilbert term considered in section \ref{sec:GR} earlier plus a k-essence term.  Thus, since we are considering no torsion, the connection could be uniquely solved as
\eqn{
\Gamma^\lambda_{~\mu\nu} \=\l\{\substack{\lambda\\\mu\nu}\r\}_{\tilde g}. \label{LCoff}
}
Substituting the solution, we obtain the Einstein-frame action written purely with the conformal metric ${\tilde g}_{\mu\nu}$ as,
\eqn{
\l.S_4^{~\text{Einstein}}\r|_{\Gamma=\{\}_{\tilde g}}=\int d^4\tilde x \sqrt{-{\tilde g}}\l[{\tilde g}^{\mu\nu}\overset{{\tilde g}}{R}_{\mu\nu}+\tilde K(\phi,\tilde X)\r],
}
where $\overset{{\tilde g}}{R}_{\mu\nu}$ is the Ricci tensor purely defined with the Levi-Civita connection of the new metric ${\tilde g}_{\mu\nu}$; we have also defined
\eqn{
\tilde K(\phi,\tilde X)=\frac{K(\phi,X)}{{G_4}^2(\phi,X)}.
}

The 'Einstein' frame of the action (\ref{eq:L4}) is none other than a minimally coupled k-essence action, which is also true for the metric formalism. Note that, however, the k-essence term for the Einstein frame differs between metric and Palatini formalisms.

Now let us consider cosmological perturbation of this Einstein frame action. The ansatz of the Einstein-frame metric and coordinates is taken as
\eqn{
d\tilde s^2=-\tilde N^2d\tilde t^2+\tilde\gamma_{ij}\l(d\tilde x^i+\tilde
N^id\tilde t\r)\l(d\tilde x^j+\tilde N^j d\tilde t\r), 
}
with $\tilde N$ and $\tilde N^i$ being the lapse and shift, respectively.
The scalar field can also be decomposed into the background quantity and the perturbation as
\eqn{
\phi(\tilde{t},\tilde{x}^i) = \phi_{\rm B}(\tilde{t}) + \delta\phi(\tilde{t},\tilde{x}^i).
}

The tensor perturbations may be calculated from the following substitution:
\eqn{
\tilde N\=1,\label{eq:tenpert1}\\
\tilde N_{i}\=0,\\
\tilde\gamma_{ij}\={\tilde a(\tilde{t})^2} \l(\delta_{ij}+\tilde h_{ij}+\frac12\delta^{kl}\tilde h_{ik}\tilde h_{jl}\r),\\
{\delta\phi} \= 0.
\label{eq:tenpert2}
}
This ansatz of the metric results in the quadratic action of the tensor perturbations in the Einstein frame as~\cite{Kobayashi:2011nu}
\eqn{
\delta^{(2)} S_{4}^{~\text{Einstein,tensor}}=\frac14\int d\tilde t\,d^3 \tilde x\,\tilde a^3\l[\tilde h^{\prime2}_{ij}-\frac{1}{\tilde a^2}(\tilde\partial_k \tilde h_{ij})^2\r],\label{eq:einsteingw}
}
where 
{the prime represents the derivative with respect to $\tilde t$}
and $\delta^{(2)}$ implies quadratic in perturbation.

Therefore, the sound speed of tensor perturbations, i.e.\ the velocity of
the gravitational waves, is unity. This is in sharp contrast with
{the case of metric formalism where, in general, the sound speed
differs from unity due to the $X$ dependence in $G_4$ and the associated
counter term in Eq.~(\ref{eq:L4metric})~\cite{Kobayashi:2011nu}.}

Similarly, let us consider the scalar perturbations. The metric and the curvature perturbations are taken as,
\eqn{
\tilde N\=1+\tilde\alpha,\label{eq:scapert1}\\
\tilde N_i\=\tilde\partial_i\tilde\beta,\\
\tilde\gamma_{ij}\=\tilde a(\tilde t)^2e^{2\tilde\psi}\delta_{ij},\\
\tilde{\zeta} \= - \tilde{\psi} + \frac{\tilde{H}(\tilde t)}{\phi'_{\rm B}(\tilde{t})}\,\delta\phi. 
\label{eq:scapert2}
}
Here $\tilde{H}=\frac{{\tilde a}^\prime}{\tilde a}$ is the Hubble parameter of the Einstein frame.
This results in the quadratic action of curvature perturbation as~\cite{Kobayashi:2011nu},
\eqn{
\delta^{(2)} S_{4}^{~\text{Einstein,scalar}}=\int d\tilde t\, d^3 \tilde x\,\tilde a^3\l[\tilde{\mathcal G_S}\zeta^{\prime2}-\frac{\tilde{\mathcal F_S}}{\tilde a^2}(\tilde\partial_i \tilde\zeta)^2\r],
}
with
\eqn{
\tilde{\mathcal F_S}\=\frac{6\tilde XK_{\tilde X}}{-\tilde K+2\tilde X\tilde K_{\tilde X}}=2\tilde\epsilon\nonumber\\
\=\frac{6X(2KG_{4X}-K_XG_4)}{(K-2XK_X)G_4+3KG_{4X}X},
\label{eq:FS}\\
\tilde{\mathcal G_S}\=\frac{6(\tilde X\tilde K_{\tilde X}+2\tilde X^2\tilde K_{\tilde X\tilde X})}{-\tilde K+2\tilde X\tilde K_{\tilde X}}\nonumber\\
\=\frac{6X}{(G_4-G_{4X}X)^2\{-K(G_4+3G_{4X})+2XK_XG_4\}}\nonumber\\
&&\times\l[-6X^2KG_{4X}^3+X(8K+5K_XX)G_4G_{4X}^2+(K_X+2K_{XX}X)G_4^3\r.\nonumber\\
&&\quad\l.-2\{K(G_{4X}+2G_{4XX}X)+XK_X(3G_{4X}-XG_{4XX})+X^2K_{XX}G_{4X}\}G_4^2\r],\nonumber\\
\label{eq:GS}
}
where $\tilde \epsilon:=-\frac{\tilde H^{\prime}}{\tilde H^2}$ and all of the quantities on the right hand sides should be understood as background ones.
Here the constraint equations on lapse and shift were solved and substituted. 

To avoid ghost and gradient instabilities, we must impose
\eqm{
\tilde{\mathcal F_S}=2\tilde{\epsilon}&>&0,\\
\tilde{\mathcal G_S}&>&0.
}
Note that this implies that the null-energy condition must be satisfied in the Einstein frame.

The sound speed of the curvature perturbations will thus be
\eqn{
\tilde c_S^2\=\frac{2\tilde{\epsilon}}{\tilde{\mathcal G_S}}
\nonumber\\
\=\frac{\tilde K_{\tilde X}}{\tilde K_{\tilde X}+2\tilde X\tilde K_{\tilde X\tilde X}}\nonumber\\
\=\l(G_4-G_{4X}X\r)^2\l(-2KG_{4X}+K_XG_4\r)\nn\\
&&\times\Big[-6X^2K G_{4X}^3+X(8K+5XK_X)G_4G_{4X}^2+(K_X+2XK_{XX})G_4^3\nn\\
&&\quad-2[K(G_{4X}+2XG_{4XX})+XK_X(3G_{4X}-XG_{4XX})+X^2K_{XX}G_{4X})]G_4^2\Big]^{-1}\nn,\\ \label{eq:cS}
}
{where all of the quantities on the right hand side should be understood as background ones.}

Under the assumption $G_4 =G_4(\phi)$, the sound speed of curvature
perturbation reduces to \eqn{ \tilde
c_S^2=\frac{K_{X}}{K_{X}+2XK_{XX}}, } which is the same as that of
k-essence in the metric formalism. Interestingly, the form of the
conformal coupling $G_4(\phi)$ does not affect the form of the
sound speed in such case.

Following~\cite{Kobayashi:2011nu}, let us first define the variables
\eqn{
\tilde f_S:=\frac{\tilde{\epsilon}^\prime}{\tilde H\tilde\epsilon},\quad \tilde g_S:=\frac{\tilde {\mathcal G}^\prime_S}{\tilde H\tilde{\mathcal G}_S}.
}
Under the assumption $\tilde\epsilon\sim\text{const.}$, $\tilde f_S\sim\text{const.}$, and $\tilde g_S\sim\text{const.}$, the power spectrum and the spectral index of the scalar perturbation are given as
\eqn{
\tilde{\mathcal P}_{\tilde\zeta} \=\l.\frac{\tilde\gamma_S}4\frac{1}{\tilde\epsilon\tilde c_S }\frac{\tilde H^2}{4\pi^2}\r|_{-\tilde k\tilde y_S=1},\\
\tilde n_S-1\=3-\tilde\nu_S,
}
where,
\eqn{
d\tilde y_S&:=&\frac{\tilde c_S}{\tilde a}d\tilde t,\\
\tilde\nu_S&:=&\frac{3-\tilde\epsilon +\tilde{g}_S}{2-2\tilde\epsilon-\tilde{f}_S+\tilde{g}_S},\\
\tilde\gamma_S&:=&2^{2\tilde\nu_S-3}\l|\frac{\Gamma(\tilde\nu_S)}{\Gamma(\frac32)}\r|^2\l(1-\tilde\epsilon-\frac{\tilde f_S}2+\frac{\tilde g_S}2\r).
}
Here $\Gamma(s)$ is understood as the Gamma function.

Furthermore, in the limit $\epsilon, f_S, g_S\ll1$, the tensor-to-scalar ratio is computed as,
\eqn{
\tilde r=16\tilde \epsilon \tilde c_S.
}
Thus all the relevant observational variables are computed. 

\subsection{Analysis in Jordan frame}

Let us return to the original Lagrangian ,
\begin{equation}
S_4^{~\text{Jordan}}= \int d^4 x \sqrt{-g} \left[G_4(\phi,X) \overset{\Gamma}{R} + K(\phi,X)  \right].
\end{equation} By directly solving the equation of motion for the connection, the solution can be obtained as,
\eqn{
\Gamma^\lambda_{~\mu\nu}=\l\{\substack{\lambda\\\mu\nu}\r\}_g+\frac12g^{\lambda\sigma}\l(2g_{\sigma(\mu}\partial_{\nu)} \ln G_4-g_{\mu\nu}\partial_\sigma\ln G_4\r)\label{eq:L4GamSol},
}
which indeed coincides with the solution that was obtained in the Einstein frame (\ref{LCoff}). 
 Notice that when $\partial_{\mu} \phi=0$ the connection reduces to
 that of Levi-Civita, and $G_4$ becomes effectively Planck mass. As a
 result, when the dynamics of the scalar ends, such as after inflation,
 the theory becomes that of Einstein practically.
 
 First, let us consider the relation of coordinates and scale factors in both frames from the background conformal transformation, $d\tilde{s}^2 = G_{4}(t) ds^2$,
 \eqn{
 - d\tilde{t}^2 + \tilde{a}(\tilde{t})^2 \delta_{ij} d\tilde{x}^i d\tilde{x}^j = - G_{4}(t) dt^2 + a(t)^2 G_{4}(t)\delta_{ij} dx^i dx^j,
 }
 from which we have the following relations:
 \eqn{
 d\tilde t\=\sqrt{G_{4}(t)}\,dt,\\
 d\tilde x\=dx,\\
 \tilde{a}(\tilde t) \= \sqrt{G_{4}(t)}\,a(t).
 }
Here $G_{4}(t)$ represents the background quantity of $G_4(\phi,X)$.

{Now}, let us calculate the tensor perturbations. The Jordan frame metric is perturbed as (\ref{eq:tenpert1})-(\ref{eq:tenpert2}) of tilde-less quantities, whereas the connection is perturbed around the background solution (\ref{eq:L4GamSol}) as
\eqn{
\Gamma^0_{~00}\=\frac12\frac{d}{dt}\ln G_4,\label{eq:Ctenpert1}\\
\Gamma^i_{~00}\=\Gamma^0_{~i0}=0,\\
\Gamma^0_{~ij}\=a^2\delta_{ij}\l(H+\frac12\frac{d}{dt}\ln G_4\r)+D_{1,ij},\\
\Gamma^i_{~j0}\=\delta^i_j\l(H+\frac12\frac{d}{dt}\ln G_4\r)+D^{~i}_{2,j},\\
\Gamma^i_{~jk}\=\partial^iD_{3jk}+2\partial_{(j}D_{4,k)}^{~i},\label{eq:Ctenpert2}
}
where $\frac{d}{dt}\ln G_4=\frac{\partial}{\partial t}\ln G_4$.
Here $D_{s,ij}$ are tensor perturbations of the connection.

After substituting this ansatz and then solving the constraint equations, one obtain the following solutions for tensor perturbation of the connection:
\eqn{
D_2^{~ij}\=\frac12\dot h^{ij},\\
 D_1^{~ij}\=a^2\l[\frac12\dot{h}^{ij}+h^{ij}\l(H+\frac12\frac{d}{dt}\ln G_4\r)\r],\\
  D^{~ij}_4\=\frac12h^{ij},\\
 D^{~ij}_3\=\frac12h^{ij}.
}
Notice that this coincides with the perturbation of the solution of the background connection (\ref{eq:L4GamSol}).\footnote{
In tensor notation,
\eqm{
\delta \Gamma^\lambda{}_{\mu\nu}
&=&\frac12\bar g^{\lambda\sigma}\l(2\nabla_{(\mu}\delta g_{\nu)\sigma}-\nabla_{\sigma}\delta g_{\mu\nu}\r)+\frac1{2}\l( \bar g_{\mu\nu}\delta g^{\lambda\sigma}-\bar g^{\lambda\sigma}\delta g_{\mu\nu}\r)\partial_\sigma \ln G_4\nn\\
&&+\frac1{2G_4}\l(\delta^\lambda_{(\mu}\delta^\sigma_{\nu)} -\frac12\bar g^{\lambda\sigma}\bar g_{\mu\nu}\r)\\
&&\quad\times\l(G_{4X}\phi_\alpha \phi_\beta\nabla_\sigma\delta g^{\alpha\beta}+2G_{4X}\phi_{\sigma(\alpha}\phi_{\beta)}\delta g^{\alpha\beta}+\phi_\alpha \phi_\beta \delta g^{\alpha\beta}\nabla_\sigma G_{4X}-G_{4X}\phi_\alpha \phi_\beta \delta g^{\alpha\beta}\nabla_\sigma \ln G_4\r)\nn,\\
}
where the metric and connection was perturbed as
$g_{\mu\nu}\to \bar g_{\mu\nu}+\delta g_{\mu\nu}$ and $\Gamma^\lambda{}_{\mu\nu} \to\bar \Gamma^\lambda{}_{\mu\nu} +\delta\Gamma^\lambda{}_{\mu\nu}$.  This could also be obtained by perturbing the $S^{~\text{Jordan}}_4$ action (\ref{eq:L4}) and solving it with respect to $\delta\Gamma^\lambda{}_{\mu\nu}$.
} 
Therefore, the tensor perturbations of the connection (\ref{eq:Ctenpert1})-(\ref{eq:Ctenpert2}) is indeed consistent.

Substituting the solutions, we obtain the quadratic action for the tensor-perturbations,
\eqn{
\delta^{(2)} S_{4}^{~\text{Jordan,tensor}}\=\delta^{(2)}\l\{\int d^4x\sqrt{-g}\l(G_4\overset\Gamma R+K\r)\r\},\nn\\
\=\frac14\int dt d^3x G_4a^3\l[\dot h_{ij}^2-\frac1{a^2}(\partial_kh_{ij})^2\r].
}
Therefore, the sound speed of tensor perturbation is indeed unity, which coincides with the analysis of the earlier section in the Einstein frame. 

{From the conformal transformation for tensor perturbed metrics, $d\tilde{s}^2 = G_{4}(t) ds^2$, we can easily verify that $\tilde h_{ij} =h_{ij}$. Thus,} the quadratic action is precisely that of (\ref{eq:einsteingw}), with the conformal redefinition of 
\eqm{
d\tilde t\={\sqrt{G_{4}(t)}}\,dt,\\
d\tilde x\=dx,\\
\tilde a(\tilde t)\={\sqrt{G_{4}(t)}}\,a(t),\\
\tilde h_{ij} \=h_{ij}.
}

As for the scalar perturbation, consider the substitution of the Jordan-frame metric and curvature perturbations (\ref{eq:scapert1})--(\ref{eq:scapert2}) as well as the following ansatz for the connection:
\eqn{
\Gamma^0{}_{00}\=\frac12\frac{d}{dt}\ln G_4+c_1,\\
\Gamma^0{}_{i0}\=c_{2,i},\\
\Gamma^0{}_{ij}\=a^2\delta_{ij}\l(H+\frac12\frac{d}{dt}\ln G_4\r)+c_3\delta_{ij}+c_{4,ij},\\
\Gamma^i{}_{00}\=c_{5}{}^{,i}\\
\Gamma^i{}_{j0}\=\delta^i_j\l(H+\frac12\frac{d}{dt}\ln G_4\r)+c_6\delta^i_j +c_7{}^{,i}{}_{,j},\\
\Gamma^i{}_{jk}\=\delta_{jk} c_8{}^{,i}+\delta^i_{(j}c_{9,k)}+c_{10}{}^{,i}{}_{,jk}.
}
Similar to the case of tensor perturbations, one may obtain the constraint equations for the auxiliary fields and solve them with respect to the curvature perturbation as
\eqn{
c_1\=\l(1+\frac{G_{4X}\dot\phi^2}{2G_4}\r)\dot\alpha+\frac{\alpha G_{4X}}{2G_4}\l[2\dot\phi \ddot\phi +\dot\phi^2\frac{d}{dt} \ln \l(\frac{G_{4X}}{G_4}\r) \r],\\
c_2\=\alpha+H\beta+\frac12\beta\frac{d}{dt}\ln G_4-\frac{G_{4X}\dot\phi^2}{2G_4}(\alpha+2H\beta),\\
c_3\=a^2\l[\dot\zeta+\l(2H+\frac{d}{dt}\ln G_4\r) (\zeta-\alpha)+\frac{G_{4X}\dot\phi^2\alpha}{2G_4}\l\{\frac{\dot\alpha}{\alpha} +2\frac{\ddot\phi}\phi+\frac{d}{dt}\ln\l(\frac{G_{4X}}{G_4}\r)\r\}\r],\\
c_4\=-\beta,\\
c_5\=\frac{1}{a^2}\l[\alpha+\dot\beta+\frac12\beta\frac{d}{dt}\ln G_4+\frac{G_{4X}\dot\phi^2}{2G_4}(\alpha+2H\beta)\r],\\
c_6\=\dot\zeta+\frac{G_{4X}\dot\phi^2\alpha}{2G_4}\l\{\frac{\dot\alpha}{\alpha} +2\frac{\ddot\phi}\phi+\frac{d}{dt}\ln\l(\frac{G_{4X}}{G_4}\r)\r\},\\
c_7\=0,\\
c_8\=-(\zeta+H\beta)+\frac{G_{4X}\dot\phi^2}{2G_4}(\alpha+2H\beta),\\
c_9\=2\zeta-\frac{G_{4X}\dot\phi^2}{G_4}(\alpha+2H\beta),\\
c_{10}\=0.
}
Finally, substituting the solutions, we obtain the quadratic action for the scalar-perturbations as
\eqn{
\delta^{(2)} S_{4}^{~\text{Jordan,scalar}}=\delta^{(2)}\l(\int d^4 x\sqrt{-g}G_4\overset\Gamma R\r)\=\int dt\,d^3x\,a^3\l[\mathcal G_S\dot \zeta^2-\frac{\mathcal F_S}{a^2}(\partial_i \zeta)^2\r].\label{eq:jordansca}
}
Here $\mathcal F_S=G_4(t)\tilde{\mathcal F}_S$ and $\mathcal G_S=G_4(t)\tilde{\mathcal F}_S$ while $\tilde{\mathcal F}_S$ and $\tilde{\mathcal G}_S$ being precisely that of ($\ref{eq:FS}$) and ($\ref{eq:GS}$) and thus the sound speed $c_S$ coinciding with (\ref{eq:cS}). 
{From the conformal transformation for the scalar-perturbed metrics, $d\tilde{s}^2 = G_{4}(t,x^i) ds^2$, we can easily verify the following relation~\cite{Chiba:2013mha}:
\eqn{
\tilde{N} \= N + \frac{2\delta G_4}{2G_{4}(t)},\\
\tilde{N}_i \= G_{4}(t) N_i.\\
\tilde{\psi} \= \psi +  \frac{2\delta G_4}{2G_{4}(t)},\\
\tilde{\zeta} \= \zeta,
}
where $\delta G_4(t,x^i) \equiv G_4(t,x^i) - G_{4}(t)$.
Thus,} the quadratic action is again the same with the one of Einstein frame up to the conformal redefinition of
\eqm{
d\tilde t\=\sqrt{G_4(t)}\,dt,\\
d\tilde x\=dx,\\
\tilde a(\tilde t)\=\sqrt{G_4(t)} \,a(t),\\
\tilde \zeta \=\zeta.
}

Since the quadratic action of tensor and scalar perturbations of the
Jordan frame are precisely that of the Einstein frame, the observables
are the same~\cite{Chiba:2013mha}. Thus, the amplitudes and the spectral index of scalar and tensor perturbations are the same. Especially, the tensor-to-scalar ratio is explicitly computed as
\eqn{
r= 16\,\frac{\mathcal F_S}{2 G_4}\,c_s = 16 \tilde{\epsilon} \tilde{c}_s=\tilde{r}.
}

\subsection{Analysis in Riemann frame}

Instead of considering the Einstein or Jordan frame, one may transform the connection and analyze in a frame where the dynamics of the metric and scalar are equivalent to that in the Jordan frame but written instead in Riemann geometry i.e.\ the connection is fixed as Levi-Civita. We shall call this frame the Riemann frame.

First noticing that the connection is not dynamical, one may substitute the solution of the connection (\ref{eq:L4GamSol}) to the action(\ref{eq:L4}) \footnote{One may also integrate out the metric, if solvable, which instead of a metric theory will become purely affine theory. Such formalism is called Eddington formalism and recently some models of inflation are considered, for example, in Ref.~\cite{Azri:2017uor,Azri:2018qux,Azri:2019ffj}.}. This results in
\eqn{
\mathcal L_4^{~\text{Riemann}}\=\sqrt{-g}\l[G_4 \overset{g}R+\frac32\frac{(\overset{g}\nabla G_{4})^2}{G_4}+K\r]\nonumber\\
\=\sqrt{-g}\l[G_4 \overset{g}R-\frac{3}{2G_4}\l(2G_{4\phi}^2X+2G_{4\phi}G_{4X}\phi^\alpha\phi_{\alpha\beta}\phi^\beta-G^2_{4X}\phi^\alpha\phi_{\alpha\beta}\phi^{\beta\gamma}\phi_\gamma\r)+K\r].\nonumber\\ \label{eq:L4ER}
}
This action is dynamically equivalent to (\ref{eq:L4}). In other words, the metric and the scalar follows the same equation of motion for both Lagrangians (\ref{eq:L4}) and (\ref{eq:L4ER}).

Interestingly, this action is a qDHOST of class ${^2N}$-I/Ia \cite{Crisostomi:2016czh,Achour:2016rkg}, as first noticed in \cite{Aoki:2018lwx}.  The cosmological perturbations of such qDHOST theories were done in \cite{Achour:2016rkg,Langlois:2017mxy,Boumaza:2020klg}, which here we also follow. 

{By defining the following functions,
\eqm{
f\=G_4,\\
\bar K\=K-\frac{3G^2_{4\phi}X}{G_4}{+}6X\frac{\partial}{\partial\phi}\int\frac{G_{4\phi}G_{4X}}{G_4} dX,\\
Q\=-3\int\frac{G_{4\phi}G_{4X}}{G_4} dX,\\
A_4\=\frac{3G_{4X}^2}{2G_4},
}
the} action (\ref{eq:L4ER}) reduces to the DHOST action considered in \cite{Achour:2016rkg,Langlois:2017mxy,Boumaza:2020klg} as \footnote{Note that the papers \cite{Achour:2016rkg,Langlois:2017mxy,Boumaza:2020klg} considered a different notation from ours, namely, they took $X=(\partial\phi)^2$. }
\eqn{
\mathcal L= f R^g+\bar K+Q\square^g\phi +A_4\phi^\mu \phi_{\mu\nu}\phi^{\nu\rho}\phi_\rho.\label{eq:DHOST}
}

The quadratic ADM action  for this action is \cite{Achour:2016rkg,Langlois:2017mxy,Boumaza:2020klg},
\eqn{
\delta^{(2)}\mathcal L^{~\text{Riemann}}_4\=a^3\frac{M^2}{2}\l\{\delta \mathcal K_{ij}\delta \mathcal K^{ij}-\delta \mathcal K^2 +\overset3{R}\frac{\delta\sqrt h}{a^3}+\delta^{(2)}\overset3R+\alpha_KH^2\delta N^2+4H\alpha_B\delta \mathcal K\delta N\r.\nonumber\\
&&\l.\qquad+(1+\alpha_ H)\overset3{R}\delta N+4\beta_1\delta \mathcal K\delta \dot N+\beta_2 \delta \dot N^2+\frac{\beta_3}{a^2}(\partial_i \delta N)^2\r\}.
}
where, $\overset3{R}$ is the spatial Ricci scalar, $\mathcal K_{ij}$ is the extrinsic curvature and
\eqm{
\frac{M^2}2\=f,\\
2HM^2\alpha_B\=-4X\dot XA_4+2(3\dot X-4HX)f_X+2\sqrt{-2X}XQ_{1X}+4X\dot X{\bar K}_{XX},\\
\frac{M^2}2H^2\alpha_K\=(24HX\dot X-3X^2+12X\ddot X)A_4+\frac12X(12HX\dot X+7\dot X^2+4X\ddot X)A_{4X}\nonumber,\\
&&+2X^2\dot X^2A_{4XX}+6X(2H^2+3\dot H)f_X+12X^2(\dot H+2H^2)f_{XX}+4X^2Q_{1X}\nonumber\\
&&-6HX^2\sqrt{-2X}Q_{1XX},\\
\alpha_H\=-2\frac{f_XX}f,\\
\alpha_T\=\alpha_L=0,\\
\beta_1\=\frac{f_XX}f,\\
\beta_2\=-6\l(\frac{f_XX}f\r)^2,\\
\beta_3\=-\frac{2f_XX}f\l(2-\frac{3f_XX}f\r).
}

Under the substitution of scalar perturbations of (\ref{eq:scapert1})-(\ref{eq:scapert2}) 
and using $\beta_2=-6\beta^2_1$, the quadratic action becomes,
\eqn{
\delta^{(2)}\mathcal L^{~\text{Riemann}}_4\=a^3\frac{M^2}2\l\{-6(\dot \zeta-\beta_1\dot \alpha)^2+12H\left[(1+\alpha_B)\dot\zeta-\beta_1\dot \alpha\right]\alpha\r.\nonumber\\
&&\qquad H^2(\alpha_K-6-12\alpha_B)\alpha^2+4\l[\dot\zeta-\beta_1\dot\alpha-H(1+\alpha_B)\alpha\r]\beta\nonumber\\
&&\qquad \l.\frac1{a^2}\l[2(\partial\zeta)^2+4(1+\alpha_H\partial_i\zeta\partial_i\alpha+\beta_3(\partial_i\alpha)^2)\r]\r\}.
}
Variation of $\beta$ gives
\eqn{
\alpha=\frac{\dot {\tilde\zeta}}{H(1+\alpha_B-\dot\beta_1)},
}
where $\tilde\zeta$ is the redefined variable of
\eqn{
\tilde \zeta=\zeta-\beta_1\alpha,
}
This results in the quadratic action of
\eqn{
\delta^{(2)}S^{~\text{Riemann}}_4=\int d^2 xdta^3\frac{M^2}2\l[A_{\tilde\zeta}\dot{\tilde\zeta}^2+B_{\tilde\zeta}\frac{\l(\partial_i{\tilde\zeta}\r)^2}{a^2}\r],
}
with
\eqn{
A_{\tilde\zeta}\=\frac1{\l(1+\alpha_B-\frac{\dot \beta_1}H\r)^2}\l[\alpha_K+6\alpha_B^2-\frac6{a^3H^2M^2}\frac{d}{dt}(a^3HM^2\alpha_B\beta_1)\r],\label{eq:Azeta}\\
B_{\tilde\zeta}\=2-\frac2{aM^2}\frac{d}{dt}\l[\frac{aM^2(1+\alpha_H+\beta_1)}{H(1+\alpha_B)-\dot\beta_1)}\r].\label{eq:Bzeta}
}

After some lengthy computation, we see that this indeed coincides with (\ref{eq:FS}) and (\ref{eq:GS}), i.e. $\frac{M^2}2A_{\tilde\zeta} =\mathcal G_S$ and $\frac{M^2}2B_{\tilde\zeta}=\mathcal F_S$.  Furthermore, we see that indeed the computation done in the Riemann frame is precisely that in the Jordan frame since the quadratic action above is precisely that of the Jordan frame~(\ref{eq:jordansca}).

To conclude, we have seen that the calculations in all frames, namely Einstein, Jordan, and Riemann frames, give the same quadratic actions, powerspectra, and sound velocities for both the scalar and tensor perturbations. In the literature, perturbations in the Einstein frame are heavily investigated, due to it being simple and straightforward. However, one may wonder what could be said for theories that do not have an Einstein frame. This we will investigate in the next section.

\section{Possible Galileon terms in Palatini formalism} 
\label{sec:Galileon}
In this section we will re-think the covariantization of the flat space-time action
\eqn{
\mathcal L_{3}^{~\text{flat}}=G_3\eta^{\mu\nu}\partial_\mu \partial_\nu\phi,
\label{eq:flatGal}}
with $G_3 =G_3\l(\phi,-\frac12\eta^{\mu\nu}\partial_\mu \phi \partial_\nu \phi\r)$.

In the usual metric formalism, the covariant action is straightforwardly obtained and unique,  which is,
\eqn{
\mathcal L_{3}^{~\text{metric}}=G_3 g^{\mu\nu}\overset{g}{\nabla}_\mu \overset{g}{\nabla}_\nu\phi.
}

However, in Palatini formalism, due to the metric incompatibility of the connection, one need to consider additional terms such as
\eqn{
\mathcal L^{~\text{Palatini}}_{3}=\l\{\begin{array}{c}G_3g^{\mu\nu}\overset{\Gamma}{\nabla}_\mu\overset{\Gamma}{\nabla}_\nu \phi\\
G_3\overset{\Gamma}{\nabla}_\mu(g^{\mu\nu}\overset{\Gamma}{\nabla}_\nu \phi)\\
G_3\overset{\Gamma}{\nabla}_\mu\{\overset{\Gamma}{\nabla}_\nu(g^{\mu\nu} \phi)\}\\
G_3g^{\mu\nu}g^{\alpha\beta}\overset{\Gamma}{\nabla}_\mu(g_{\alpha\beta}\overset{\Gamma}{\nabla}_\nu \phi)\\
\vdots\end{array}\r.,
}
One may wonder how many possible terms there could emerge, or even if it is finite at all.  To write down the possible terms, one must note the following three points.

Firstly, one notices that the covariant derivative $\overset\Gamma\nabla$ acting on any rank tensor relates to the one of Levi-Civita $\overset{g}\nabla$ as
\eqn{
\overset{\Gamma}{\nabla}=\overset{g}{\nabla}+\text{ terms containing }Q_\lambda^{~\mu\nu}~,
}
since any (torsionless) affine connection can always be rewritten as
\eqn{
\Gamma^\lambda_{~\mu\nu} \=\l\{\substack{\lambda\\\mu\nu}\r\}_g+\frac12\l(Q_{\mu\nu}^{~~\lambda}+Q_{\nu\mu}^{~~\lambda}-Q^\lambda_{~\mu\nu}\r).
}
Secondly, the covariantized terms can be schematically written as,\eqn{ G_3
\cdot \overset{\Gamma}{\nabla} \cdot \overset{\Gamma}{\nabla} \cdot \phi~,
}
with $\cdot$ representing an arbitrary number of metrics. The first derivative acts either on a metric, a covariant derivative, or $\phi$, whereas the second derivative acts on a metric or $\phi$.
Finally, the resultant terms must be a scalar, i.e., all of the space-time indices must be contracted.

With the above in hand, the covariantization  in Palatini formalism of the flat action~(\ref{eq:flatGal}) is constructed through all the possible contractions in the form of
  $\overset{g}{\nabla}\overset{g}{\nabla}\phi$, $Q\times \partial \phi$, $Q\times Q$, $\overset{g}{\nabla} Q$, which are the following 1+9 terms:
\begin{enumerate}
    \item $\overset{g}{\square}\phi$,
    \item $Q^\mu \partial_\mu \phi$, ${\tilde Q}^\mu \partial_\mu \phi$,
    \item $\phi Q_{\alpha\beta\gamma}Q^{\alpha\beta\gamma}$,  $\phi Q_{\alpha\beta\gamma} Q^{\beta\gamma\alpha}$, $\phi Q^\mu Q_\mu $, $\phi Q_\mu {\tilde Q}^\mu$, $\phi {\tilde Q}_\mu {\tilde Q}^\mu$,
     \item $\phi\overset{g}{\nabla}_\mu Q^{\mu}$, $\phi \overset{g}{\nabla}_\mu {\tilde Q}^{\mu}$,
\end{enumerate}
where we have defined the Weyl vector and the other trace vector of non-metricity as
\eqn{
Q_\mu&:=&\frac14 Q_{\mu\nu}{}^{\nu},\\
{\bar Q}^\mu&:=&Q_{\nu}{}^{\nu\mu}.
}
Therefore the most general Palatini $L_3$ action consists of 10 different terms and are given as
\eqn{
\mathcal L^{~\text{Jordan}}_3&:=&G_{3,0}\overset{g}{\square}\phi+G_{3,1}Q^\mu \partial_\mu \phi+G_{3,2}{\bar Q}^\mu \partial_\mu \phi\nonumber\\
&&+G_{3,3}\phi Q_{\alpha\beta\gamma}Q^{\alpha\beta\gamma}+G_{3,4}\phi Q_{\alpha\beta\gamma} Q^{\beta\gamma\alpha}+G_{3,5}\phi Q^\mu Q_\mu +G_{3,6}\phi Q_\mu {\bar Q}^\mu\nonumber\label{eq:L3terms}\\
&&+G_{3,7}\phi {\bar Q}_\mu {\bar Q}^\mu+G_{3,8}\phi \overset{g}{\nabla}_\mu Q^{\mu}+G_{3,9}\phi \overset{g}{\nabla}_\mu {\bar Q}^{\mu}\\
&=:&\sum_{i=0}^9G_{3,i}~\overset\Gamma\square_{(i)}\phi \label{eq:L3},
}
where the arguments of all the functions are $\phi$ and $X$, i.e.\ $G_{3,i}=G_{3,i}(\phi,X)$, etc.  Under the flat space-time limit of $g_{\mu\nu}\to \eta_{\mu\nu}$ and $\Gamma^\lambda{}_{\mu\nu}\to0$, the Palatini L3 action (\ref{eq:L3}) indeed reduces to the flat space-time action (\ref{eq:flatGal}).
  
The first three terms in (\ref{eq:L3terms}) were considered in Refs.~\cite{Shimada:2018lnm,Kozak:2018vlp,Helpin:2019kcq,Helpin:2019vrv}, whereas the first eight terms were in Ref.~\cite{Rasanen:2018ihz} with the functions $G_{3,i}$ only being dependent on the scalar. The most general form of $L_3$ has, to our knowledge, not been investigated.

Furthermore, recall that the Riemann tensor is the form of $\overset\Gamma{R}\sim\partial\Gamma+\Gamma\Gamma$. Similarly, the terms $Q\times Q$ and $\overset{g}\nabla Q$  also inhere such structure of $\Gamma\Gamma$ and $\partial\Gamma$ respectively. One then might guess that these terms might affect the results of the cosmological perturbations significantly, such as the speed of gravitational waves. Especially, one may wonder if the speed of gravitational waves could deviate from unity, different from the case of metric formalism. We shall investigate these issues in the next session.
\section{Tensor and scalar perturbations with the Galileon terms in
 Palatini formalism}
\label{sec:quadratic}
Here, noting the previous section, we consider the following Lagrangian,
\eqn{
\mathcal L^{~\text{Jordan}}_{3+4}:=\mathcal L^{~\text{Jordan}}_3+\mathcal L^{~\text{Jordan}}_4=G_4\overset\Gamma R+K+\sum_{i=0}^9G_{3,i}~\overset\Gamma\square_{(i)}\phi\label{eq:L34},
}
where $G_4$, $K$,$G_{3,i}$ are understood to be functions of $\phi$ and $X=-\frac12g^{\mu\nu}\partial_\mu \phi \partial_\nu \phi$.

Unlike what was considered in Sec.~\ref{sec:Ricci},  the conformal transformation of this Lagrangian, due to the existence of the L3 term and its connection dependence, does not lead to the Einstein frame. Similarly, analysis in the Jordan frame will be tedious. We therefore shall resort to analysis in the Riemann frame.

The connection for this Lagrangian (\ref{eq:L34}) can be solved as,
\eqn{
\Gamma^\lambda_{~\mu\nu}=\l\{\substack{\lambda\\\mu\nu}\r\}_g+\frac1D\l[\l\{A^X\partial^\lambda X +A^\phi\partial^\lambda\phi\r\}g_{\mu\nu}+2\l\{B^X\partial_{(\mu}X+B^\phi\partial_{(\mu}\phi\r\}\delta^\lambda_{\nu)}\r],
}
where,
\eqn{
A^\phi\=-6G_4(2G_{4\phi}+G_{3,1}+2G_{3,2}-G_{3,8}-2G_{3,9})\nn\\
&&-\l\{-6(G_{3,8\phi}+2G_{3,9\phi})G_4+2[32(G_{3,3}+G_{3,4})+5G_{3,5}+17G_{3,6}+56G_{3,7}]G_{4,\phi}\r.\nn\\
&&\quad +(8G_{3,3}+12G_{3,4}+5G_{3,6}+28G_{3,7})G_{3,1}-2(16G_{3,3}+8G_{3,4}+5G_{3,5}+7G_{3,6})G_{3,2}\nn\\
&&\quad-8G_{3,3}G_{3,8}-12G_{3,4}G_{3,8}-5G_{3,6}G_{3,8}-28G_{3,7}G_{3,8}+32G_{3,3}G_{3,9}+16G_{3,4}G_{3,9}\nn\\
&&\l.\quad+10G_{3,5}G_{3,9}+14G_{3,6}G_{3,9}\r\}\phi\nn\\
&&+\{(8G_{3,3}+12G_{3,4}+5G_{3,6}+28G_{3,7})G_{3,8\phi}-2(16G_{3,3}+8G_{3,4}+5G_{3,5}+7G_{3,6})G_{3,9\phi}\}\phi^2,\nn\\
\ \\
A^X\=6G_4\{-2G_{4X}+(G_{3,8X}+2G_{3,9X})\phi\}\nn\\
&&-2\{32(G_{3,3}+G_{3,4})+5G_{3,5}+17G_{3,6}+56G_{3,7}\}G_{4X}\phi\nn\\
&&+\{(8G_{3,3}+12G_{3,4}+5G_{3,6}+28G_{3,7})G_{3,8X}-2(16G_{3,3}+8G_{3,4}+5G_{3,5}+7G_{3,6})G_{3,9X}\}\phi^2,\nn\\
}
\eqn{
B^\phi\=4G_4(6G_{4\phi}+G_{3,1}-2G_{3,2}-G_{3,8}+2G_{3,9})\nn\\
&&+2\l\{-2(G_{3,8\phi}-2G_{3,9\phi})G_4+2[16(G_{3,3}+G_{3,4})+G_{3,5}+7G_{3,6}+40G_{3,7}]G_{4,\phi}\r.\nn\\
&&+(8G_{3,3}+4G_{3,4}+G_{3,6}+20G_{3,7})G_{3,1}-2(8G_{3,4}+G_{3,5}+5G_{3,6})G_{3,2}-8G_{3,3}G_{3,8}\nn\\
&&\l.-4G_{3,4}G_{3,8}-G_{3,6}G_{3,8}-20G_{3,7}G_{3,8}+16G_{3,4}G_{3,9}+2G_{3,5}G_{3,9}+10G_{3,6}G_{3,9}\r\}\phi\nn\\
&&+2\{(8G_{3,3}+4G_{3,4}+G_{3,6}+20G_{3,7})G_{3,8\phi}+2(8G_{3,4}+G_{3,5}+5G_{3,6})G_{3,9\phi}\}\phi^2,\nn\\
\ \\
B^X\=4G_4\{6G_{4X}-(G_{3,8X}-2G_{3,9X})\phi\}+4\{16(G_{3,3}+G_{3,4})+G_{3,5}+7G_{3,6}+40G_{3,7}\}G_{4X}\phi\nn\\
&&+\{2(8G_{3,3}+4G_{3,4}+G_{3,6}+20G_{3,7})G_{3,8X}-4(8G_{3,4}+G_{3,5}+5G_{3,6})G_{3,9X}\}\phi^2,\nn\\
\ \\
D\=24G_4^2+4(8G_{3,3}+20G_{3,4}-G_{3,5}+8G_{3,6}+44G_{3,7})G_4\phi\nn\\
&&-2\l\{64G_{3,3}^2-32G_{3,4}^2+16G_{3,3}(2G_{3,4}+G_{3,5}+G_{3,6}+10G_{3,7})-9(G_{3,6}^2-4G_{3,5}G_{3,7})\nn\r.\\
&&\l.\qquad+4G_{3,4}[G_{3,5}-8(G_{3,6}+G_{3,7})]\r\}\phi^2.
}
When $G_{3,i}=0$, this indeed reduces to (\ref{eq:L4GamSol}). Again notice that under $\partial_{\mu} \phi=0$ the connection reduced to that of Levi-Civita.

Substituting the solutions of the connection to~(\ref{eq:L34}) the Riemann frame of this action after some calculation becomes
\eqn{
\mathcal L^{~\text{Riemann}}_{3+4}\=G_4\overset{g}R+K+ G_{3,0}\overset{g}\square\phi+E_{\phi\phi}+E_{\phi X}\phi^\alpha\phi_{\alpha\beta}\phi^\beta+E_{X X}\phi^\alpha\phi_{\alpha\beta}\phi^{\gamma\delta}\phi_\delta~,\label{eq:ERL34}
}
where,
\eqn{
E_{\phi\phi}\=-\frac{X}{8D^2}\l[4A^{\phi 2}\l\{12G_4+(40G_{3,3}+28G_{3,4}+G_{3,5}+10G_{3,6}+100G_{3,7})\phi\r\}\nn\r.\\
&&\qquad+B^{\phi 2}\l\{12G_4+(136G_{3,3}+124G_{3,4}+25G_{3,5}+70G_{3,6}+196G_{3,7})\phi\r\}\nn\\
&&\qquad+4A^\phi B^\phi\l\{24G_4 +(56G_{3,3}+68G_{3,4}+5G_{3,5}+32G_{3,6}+140G_{3,7})\phi\r\}\nn\\
&&\qquad-8DA^X\{6G_{4\phi}-G_{3,1}-10G_{3,2}+G_{3,8}+10G_{3,9}+(G_{3,8\phi}+10G_{3,9\phi})\phi\}\nn\\
&&\qquad\l.+4DB^X\{6G_{4\phi}+5G_{3,1}+14G_{3,2}-5G_{3,8}-14G_{3,9}-(5G_{3,8\phi}+14G_{3,9\phi})\phi\}\r],\nn\\
\ \\
E_{\phi X}\=-\frac1{8D^2}\l[A^\phi A^X\l\{48 G_4+4(40G_{3,3}+28G_{3,4}+G_{3,5}+10G_{3,6}+100G_{3,7})\phi\r\}\r.\nn\\
&&\qquad +(B^\phi A^X+A^\phi B^X)\l\{48 G_4+2(56G_{3,3}+68G_{3,4}+5G_{3,5}+32G_{3,6}+140G_{3,7})\phi\r\}\nn\\
&&\qquad +B^\phi B^X\l\{12 G_4+(136G_{3,3}+124G_{3,4}+25G_{3,5}+70G_{3,6}+196G_{3,7})\phi\r\}\nn\\
&&\qquad -4DA^\phi(6G_{4X}+G_{3,8X}\phi+10G_{3,9X}\phi)\nn\\
&&\qquad +2DB^\phi(6G_{4X}-5G_{3,8X}\phi-14G_{3,9X}\phi)\nn\\
&&\qquad -4DA^X\l\{6G_{4\phi}-G_{3,1}-10G_{3,2}+G_{3,8}+10G_{3,9}+(G_{3,8\phi}+10G_{3,9\phi})\phi\r\}\nn\\
&&\qquad\l. +2DB^X\l\{6G_{4\phi}+5G_{3,1}+14G_{3,2}-5G_{3,8}-14G_{3,9}-(5G_{3,8\phi}+14G_{3,9\phi})\phi\r\}\r],\nn\\}
\eqn{
E_{XX}\=\frac{1}{16D^2}\l[4A^{X2}\l\{12G_4+(40G_{3,3}+28G_{3,4}+G_{3,5}+10G_{3,6}+100G_{3,7})\phi\r\}\nn\r.\\
&&\qquad+B^{X2}\l\{12G_4+(136G_{3,3}+124G_{3,4}+25G_{3,5}+70G_{3,6}+196G_{3,7})\phi\r\}\nn\\
&&\qquad+4A^XB^X\l\{24G_4 +(56G_{3,3}+68G_{3,4}+5G_{3,5}+32G_{3,6}+140G_{3,7})\phi\r\}\nn\\
&&\qquad-8DA^X\{6G_{4X}+(G_{3,8X}+10G_{3,9X})\phi\}\nn\\
&&\qquad\l.+4DB^X\{6G_{4X}-(5G_{3,8X}+14G_{3,9X})\phi\}\r].
}
Again, indeed under $G_{3,i}=0$, this reduces to the action (\ref{eq:L4ER}).

Unlike the action (\ref{eq:L4}) we previously considered, this action, in general, will have ghost degrees of freedom, namely the Ostrogradsky instability.  To eliminate this Ostrogradsky ghost, one must impose the condition
\eqn{
E_{XX}=\frac{3G^2_{4X}}{2G_4}. \label{eq:degen}
}
As a result, the theory (\ref{eq:L34}) will have at most 2 tensor and 1 scalar degrees of freedom. This again falls into the qDHOST class of $^2$N-I/Ia~ \cite{Crisostomi:2016czh,Achour:2016rkg}.
Thus, similar to Sec.~\ref{sec:Ricci} the tensor perturbation of this theory, under the condition (\ref{eq:degen}), has the sound velocity of unity, which coincides with that in metric formalism. Thus, contrary to the naive expectation, once one removes the ghost degree of freedom, the L3 terms in Palatini formalism can still serve as dark energy as well as an inflaton.

Furthermore, again, under the redefinition of functions
\eqn{
f\=G_4,\\
\bar K\=K +E_{\phi\phi}+2X\frac{\partial}{\partial\phi}\int E_{\phi X}dX,\\
Q\=G_{3,0}+\int E_{\phi X}dX,\\
A_4\=\frac{3G^2_{4X}}{2G_4}=E_{XX},
}
the action (\ref{eq:ERL34}) reduces to the DHOST action (\ref{eq:DHOST}) which can then be used to estimate the scalar perturbations. Due to tediousness, we shall omit the explicit form of the sound speed of scalar perturbations, however it can be computed from (\ref{eq:Azeta}) and (\ref{eq:Bzeta}) following the lines of Sec.~3.3.

\section{Conclusion and discussions}
\label{sec:con}
 
In this paper, we considered the cosmological perturbations of scalar-tensor theories in the Palatini formalism. First of all, we discuss the action (\ref{eq:L4}), where the Ricci scalar is conformally coupled to a function of a scalar and its kinetic term, and there is k-essence action consisting of a scalar and its kinetic term. We have found that for such a non-minimally coupled theory of (\ref{eq:L4}), there are three (classically) equivalent frames, Jordan, Einstein, and Riemann; have computed their quadratic formulae for tensor and scalar perturbations; and have shown their equivalence. Notably, the tensor modes propagate with the sound velocity of unity, which is different from the metric formalism counterpart. Thus, this model can serve as dark energy as well as an inflaton despite the presence of $X$ dependence in the $G_4$ term.

Next we considered the extension of the $\mathcal L_3$ terms called Galileon terms to the Palatini formalism as in (\ref{eq:L34}), which does not have an Einstein frame. A term corresponding to $\Box\phi$ in metric formalism is not uniquely defined in Palatini formalism due to the covariant derivative not being compatible with the metric, that is, non-metricity. We found that there are essentially 10 different (inequivalent) definitions in Palatini formalism for such a term.
By including all of them in our action, we have also computed its perturbations.
One might expect that the L3 terms can generate a ghost due to Ostrograsky instability and the sound speed of gravitational waves could potentially deviate from unity, in sharp contrast to the case of metric formalism. However, imposing the ghost-free conditions leads to the speed of the tensor modes to be unity, whereas the scalar-perturbations differ in general.
This fact is quite interesting because the ghost-free L3 terms in Palatini formalism can still serve as dark energy as well as an inflaton.
 
Similar to (\ref{sec:Galileon}), one may want to consider higher terms associated with the scalar such as $(\nabla^\Gamma\nabla^\Gamma\phi)^2$ to implement $\mathcal L_4$ terms or $\mathcal L_5$ terms. This, however, introduces the kinetic term for the connection in general. Thus the theory will exhibit more than 3 degrees of freedom. This implies that one cannot analyze neither in the Einstein frame nor in the Riemann one.  Furthermore, one can say that such theory is similar to quadratic gravity in Palatini/metric-affine formalism, which also has gained increasing interest in recent years~\cite{Exirifard:2007da,Borunda:2008kf,Vassiliev:2003dk,BeltranJimenez:2019acz,Aoki:2019snr,BeltranJimenez:2019acz,Percacci:2019hxn,Jimenez:2020dpn,Aoki:2020zqm,Aoki:2020rae}.  However, these are left for future work.
 
Finally, we would also like to comment that scalar-tensor theories in Palatini formalism are yet to be fully analyzed. Up to our knowledge, neither generalized scalar-tensor theories with a dynamical nor a non-dynamical connection that admit second order equations of motion are not known. It will be interesting to follow Lovelock's and Horndeski's footsteps to find such a theory. However, this is also left for future work.
 
\section*{Acknowledgments}\ 
K.S. will like to thank Dr. Katsuki Aoki for fruitful and very enlightening discussions. He will also like to thank the Central European Institute for Cosmology and Fundamental Physics, where part of this work was conducted, for their warm and welcoming hospitality. He also acknowledges the xTras package \cite{Nutma:2013zea} of Mathematica, which was used for the tensor calculations. K.\,O.\ is in part supported by JSPS KAKENHI Grant Number 19H01899. K.\,S.\ is supported by JSPS KAKENHI Grant Number JP20J12585. M.\,Y.\ is supported in part by JSPS Grant-in-Aid for Scientific Research Numbers 18K18764, Mitsubishi Foundation, and JSPS Bilateral Open Partnership Joint Research Projects.


\bibliography{references}
\bibliographystyle{JHEP}


\end{document}